# STATUS AND PLANS FOR AN SRF ACCELERATOR TEST FACILITY AT FERMILAB*

M. Church[#], J. Leibfritz, S. Nagaitsev, FNAL, Batavia, IL 60150, U.S.A.

*Abstract*

A superconducting RF accelerator test facility is currently under construction at Fermilab. The accelerator will consist of an electron gun, 40 MeV injector, beam acceleration section consisting of 3 TTF-type or ILC-type cryomodules, and multiple downstream beam lines for testing diagnostics and performing beam experiments. With 3 cryomodules installed this facility will initially be capable of generating an 810 MeV electron beam with ILC beam intensity. The facility can accommodate up to 6 cryomodules for a total beam energy of 1.5 GeV. This facility will be used to test SRF cryomodules under high intensity beam conditions, RF power equipment, instrumentation, and LLRF and controls systems for future SRF accelerators such as the ILC and Project-X. This paper describes the current status and overall plans for this facility.

## FACILITY DESCRIPTION

Figure 1 shows the overall layout of the new facility. The existing New Muon Lab building (NML) is 70 m in length and was previously used to house a high energy fixed target experiment. This building has been completely refurbished to accommodate the upstream end of the beamlines and supporting infrastructure. Starting from the lower left in the figure are the RF gun, injector, and 1$^{st}$ 3 SRF cryomodules, all housed in the original building. A 70 m tunnel extension has been built to house the high energy end of the beamlines and high power beam dumps. The cryogenic plant currently consists of 2 Tevatron-style satellite refrigerators and does not have the capacity to cool all 3 cryomodules at full beam intensity and at full pulse repetition rate. Also shown in Figure 1 is a new refrigeration plant (CMTF) [1] which is being built with capacity to deliver 600 W of 2 °K helium. CMTF will also have multiple horizontal cryomodule test stands to test 325 MHz, 650 MHz, and 1.3 GHz SRF cryomodules in CW mode.

### Photocathode Laser

An enclosed, temperature-controlled laser hut is being built inside the NML building to house the photocathode laser. The seed laser is a fiber optic laser (from Calmar Laser) operating at 1.3 GHz with Ytterbium as the lasing element and output at 1060 nm. A Pockels cell reduces the pulse rate to 81.25 MHz and further amplification and an additional Pockels cell reduces the rate to 3 MHz and increases the pulse energy to ~1 mJ/pulse. Two doubling crystals then decrease the wavelength to 265 nm which is used to excite the $Cs_2Te$ photocathode in the RF gun. This laser will be capable of producing 10's of nC of charge per pulse with rms pulse lengths of 2 ps – 20 ps.

In addition, there will also be a TiSa laser housed in the same laser hut capable of exciting the photocathode with pulses of rms pulse length of 100 fs.

### RF Electron Gun

The RF electron gun is identical to the guns recently developed at DESY Zeuthen (PITZ) for the FLASH facility [2]. It is a 1½ cell L-band gun capable of operating at up to 5 MW peak power and DC power dissipation of ~20 KW. We expect to operate the gun routinely at peak gradients of 45 MV/m, and output kinetic energy of ~5 MeV. The photocathode is $Cs_2Te$ with 5 mm diameter sensitive area. These photocathodes are coated at a separate facility on the Fermilab site and transferred to the gun under vacuum. The photocathode preparation facility was developed by D. Sertore at INFN Milano. The gun is surrounded by 2 solenoid magnets for emittance compensation, and ASTRA simulations indicate that transverse normalized transverse emittances of 5 μm can be attained at a bunch charge of 3.2 nC. RF windows are currently being conditioned and gun commissioning will commence in late 2011.

### Booster Cavities

The electron gun is followed by two SRF cryomodules to accelerate beam to 40 – 50 MeV. Each cryomodule contains a 9-cell L-band cavity operating at 1.3 GHz. One cavity is currently in use at the Fermilab A0 Photoinjector [3] operating at a peak gradient of 12 MV/m and will be refurbished with a higher gradient cavity in the Fall of 2011 before installation at NML. The other cavity is currently installed at NML and operates at a peak gradient of 25 MV/m.

### 40 MeV Injector line

Downstream of the SRF booster cavities is the 40 MeV injector beamline whose primary purpose is to deliver appropriate beam to the SRF cryomodules. This beamline will have adequate instrumentation to fully characterize beam parameters – toroids for intensity, BPMs for position, OTR and YAG screens for beam size, slit masks for transverse emittance, spectrometer magnet for energy and energy width, phase monitors for timing jitter, and streak camera and interferometer for bunch length. There is a 4-dipole chicane for bunch compression, and there will eventually be a 3.9 GHz SRF cavity for bunch linearization. A 3.9 GHZ normal conducting transverse mode cavity [4] will be installed to measure longitudinal phase space (in conjunction with the spectrometer magnet).

___

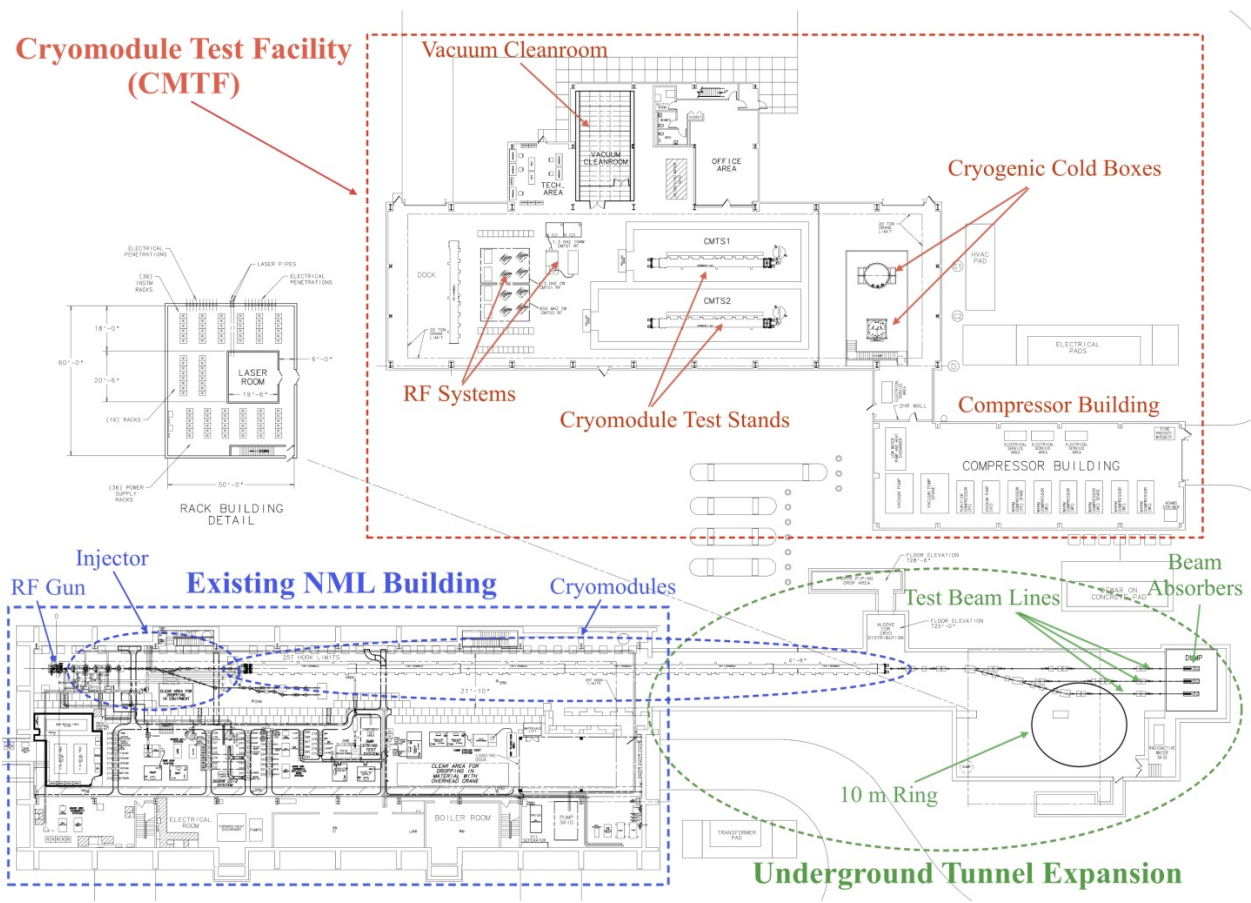

Figure 1: Overall layout of the new SRF Accelerator Test Facility at FNAL.

There is a 40 MeV dump capable of handling full beam power of 2.5 KW. There is also space for the installation of parallel 40 MeV beamlines for additional advanced accelerator R&D experiments. The beamline layout from gun to the 1st cryomodule is shown in Figure 2.

### Acceleration Section

The acceleration section of the beamline has space for up to 6 TTF-type or ILC-type SRF cryomodules. Each cryomodule consists of 8 9-cell 1.3 GHz cavities with peak gradients at least as high as 31 MV/m. A group of 3 cryomodules will be driven by a single 10 MW klystron. The first TTF-type cryomodule has been installed and is currently being tested and the results of these tests are reported elsewhere at this Workshop [5]. We intend to commence beam operations with a single TTF-type cryomodule in late 2012, operating with beam energies up to 250 MeV. In late 2014 we intend to install 2 additional ILC-type cryomodules to constitute a single ILC RF unit and will operate at beam energies up to 900 MeV. At this time the new refrigeration plant (CMTF) will be made operational and the ILC "RF unit test" (S2) can commence at full ILC beam intensity.

Proposals to add additional cryomodules beyond the initial 3 are currently being considered. There is also consideration of adding a 2nd bunch compressor, an additional cryomodule, and an undulator to produce FEL radiation in the far ultraviolet [6].

### High Energy Beamlines

In the initial stage, there will be 2 high energy beamlines terminating in 2 beam dumps capable of handling 75 KW of beam power each. In addition, there is space for a future 3rd high energy beamline and adequate floor space to accommodate a 10 m diameter test ring. There is a total of ~20 m of open beamline space to mount advanced accelerator R&D experiments in the 2 initial beamlines. There will be spectrometer magnets to measure beam energy at 2 separate locations with an accuracy of ~1 part in $10^4$. The high energy beamline layout is shown in Figure 3.

Several workshops [7] [8] have been held in the past 5 years to solicit interest and ideas for advanced accelerator R&D experiments at this facility, both at the high energy end and the low energy end. In particular, we intend to build a 10 m diameter storage ring at the high energy end to operate at 150 – 250 MeV to test the principle of nonlinear integrable optics for high intensity storage rings [9].

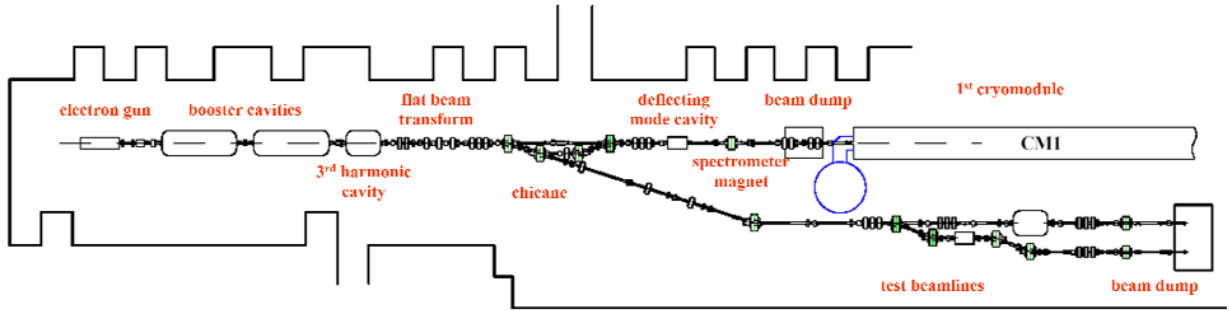

Figure 2: Layout of the injector beamlines from the RF gun to the 1st cryomodule.

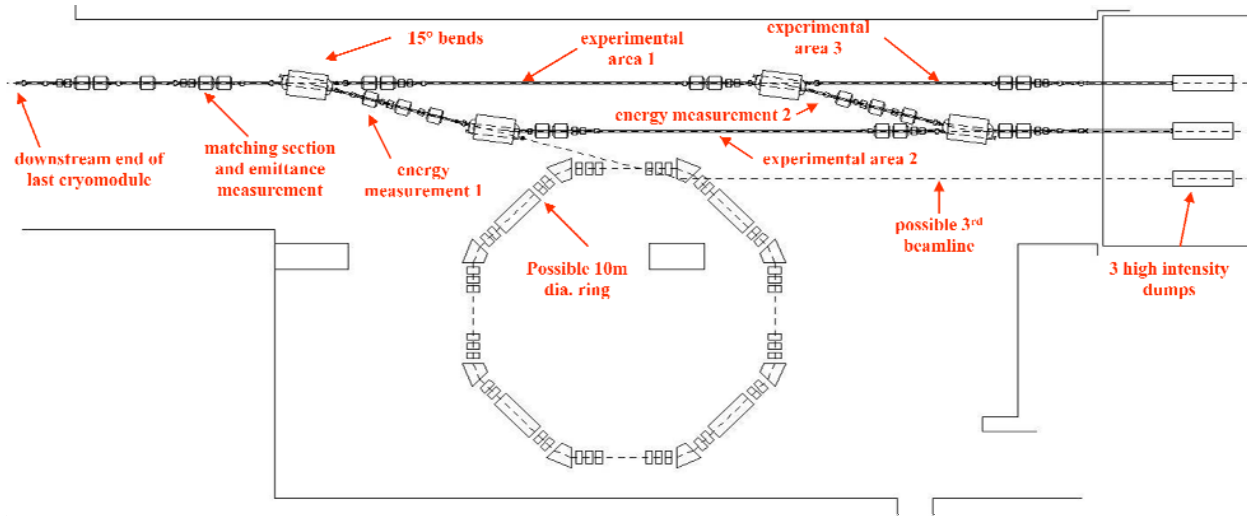

Figure 3: Layout of the high energy beamlines.

## BEAM PARAMETERS

The SRF Accelerator Test Facility will be capable of operating over a wide range of beam parameters. The machine will be operated in pulsed mode, with a pulse rate of up to 5 Hz, and each pulse up to 1 msec in length and up to 3000 bunches long. The bunch separation within a pulse is 333 nsec (3 MHz). Table 1 lists the beam parameters which will be used for the ILC RF unit test, which generally represents the maximum current the facility will handle. However, each parameter can be varied over a wide range, as long as the total beam power remains below the limit determined by the radiation shielding assessment and the beam dump capability.

Start-to-end beamline simulations have been done under a variety of conditions [10] to understand beam behaviour. Currently, the code ASTRA is used for RF gun simulation, Elegant for single particle dynamics, and IMPACT-Z for multiparticle dynamics.

Table 1: Beam Parameters

| Parameter | ILC RF unit test | range |
|---|---|---|
| bunch charge | 3.2 nC | 0.05 nC to >20 nC |
| bunch spacing | 333 nsec | <12 nsec to 0.1 sec |
| bunch train length | 1 msec | 1 bunch to 1 msec |
| bunch train repetition rate | 5 Hz | 0.1 Hz to 5 Hz |
| norm. transverse emittance | ~10 µm | ~1 µm to ~50 µm |
| RMS bunch length | 1 ps | 100 fs to 20 ps |
| peak bunch current | 3 kA | 10 kA |
| injection energy | 40 MeV | 5 MeV to 50 MeV |
| high energy | 810 MeV | 40 MeV to 1500 MeV |

## ACKNOWLEDGEMENTS

We wish to thank the entire Project Team at NML for their valuable contributions.

## REFERENCES

[1] J. Leibfritz, *et al.*, "Status and Plans for a SRF Accelerator Test Facility at Fermilab", NA-PAC'11, New York, March 2011, MOP009; http://www.JACoW.org.

[2] M. Krasilnikov, *et al.*, "Recent Developments at PITZ", PAC'05, Knoxville, May 2005, WPAP006

[3] E. Colby, "Design, Construction, and Testing of a Radiofrequency Electron Photoinjector for the Next Generation Linear Collider", Fermilab-Thesis-1997-03 (1997); http://lss.fnal.gov/archive/thesis/1900/fermilab-thesis-1997-03.pdf

[4] T. Koeth, "An Observation of Transverse to Longitudinal Emittance Exchange at the Fermilab A0 Photoinjector", Fermilab-Thesis-2009-55 (2009); http://lss.fnal.gov/archive/thesis/fermilab-thesis-2009-55.pdf

[5] E. Harms, *et al.*, "RF Test Results from Cryomodule 1 at the Fermilab SRF Beam Test Facility", SRF'11, Chicago, August 2011, MOPO013; http://www.JACoW.org.

[6] A. Lumpkin, *et al.*, "Consideration on Fermilab's Superconducting Test Linac for an EUV/Soft X-ray SASE FEL", FEL'10, Malmo, Sweden, August 2010, MOPC13; http://www.JACoW.org.

[7] Workshop on Future Directions for Accelerator R&D at Fermilab, Lake Geneva, WI, May 2009; http://apc.fnal.gov/ARDWS/index.html

[8] Workshop on Possible Directions for Advanced Accelerator R&D at the ILC Test Accelerator at Fermilab, Batavia, IL, November 2006; http://home.fnal.gov/~piot/ILCTA_AARD/

[9] A. Valishev, *et al.*, "Ring for Test of Nonlinear Integrable Optics", NA-PAC'11, New York, March 2011, WEP070; http://www.JACoW.org.

[10] C. Prokop, *et al.*, "Start-to-end Beam Dynamics Simulations for the SRF Accelerator Test Facility at Fermilab", NA-PAC'11, New York, March 2011, WEP036; http://www.JACoW.org.